\documentclass[conference]{IEEEtran}

\usepackage{cite}
\usepackage{amsmath}
\usepackage[utf8]{inputenc}
\usepackage[english]{babel}
\usepackage{graphicx}
\usepackage{algorithmic}
\usepackage{array}
\usepackage{url}

\DeclareMathOperator*{\argmax}{arg\,max}

\newtheorem{theorem}{Theorem}

\begin{document}
%
\title{Generic LSH Families for the Angular Distance Based on Johnson-Lindenstrauss Projections and Feature Hashing LSH}

\author{\IEEEauthorblockN{Luis Argerich}
\IEEEauthorblockA{Universidad Nacional de Tres de Febrero \\
y\\
Facultad de Ingeniería \\
Univesidad de Buenos Aires\\
Email: largerich@fi.uba.ar \\
}
\and
\IEEEauthorblockN{Natalia Golmar}
\IEEEauthorblockA{Facultad de Ingeniería \\
Universidad de Buenos Aires\\
Email: ngolmar@gmail.com}
}


\maketitle

\begin{abstract}
In this paper we propose the creation of generic LSH families for the angular distance based on Johnson-Lindenstrauss projections. We show that feature hashing is a valid J-L projection and propose two new LSH families based on feature hashing. These new LSH families are tested on both synthetic and real datasets with very good results and a considerable performance improvement over other LSH families. While the theoretical analysis is done for the angular distance, these families can also be used in practice for the euclidean distance with excellent results [2]. Our tests using real datasets show that the proposed LSH functions work well for the euclidean distance.
\end{abstract}

\section{Introduction}
\textit{Locality Sensitive Hashing} (LSH) is a modern solution to the Approximate Nearest Neighbors problem (ANN) for large datasets. The basic idea of LSH is to hash similar items to the same bucket. After hashing a query point we can recover the points in the bucket as candidates and compute the distance to the query only for those points. This is a significant improvement over the brute force approach of comparing the query point against all the other points in the dataset. 

In this paper we work with LSH for the angular distance over the unit sphere. The angular distance is widely used in areas such as information retrieval and word embeddings.[13][14][15]
In the practice the LSH schemes created for the angular distance can also be used for the euclidean distance with very good results. This work will propose two new LSH Families based on Feature Hashing and show, in practice, that they have similar results to other well known LSH families and significant performance improvements for both synthetic and real datasets. We will also generalize the LSH families for the angular distance to the use of any form of a Johnson-Lindenstrauss projection and show that classical LSH families for the angular distance can be generalized deriving a minhash function from a Johnson-Lindenstrauss projection.

\section{Preliminaries and Notation}

LSH was introduced by Indyk and Motwani in 1998 [7][6]. They basic idea of LSH is to hash similar points to the same bucket, this allows approximate $O(1)$ query time when retrieving near neighbors. Given a query point we hash the point and then go to the bucket pointed by the LSH function and compare the query against the points found in the bucket. A very good LSH function will minimize the number of points to be compared and maximize the number of real near neighbors found in the bucket. 

We define a \textit{minhash} $h(x) \rightarrow [0..m)$ as the result of applying a hash function to a $d$ dimensional point $x$ and obtaining a bucket number to store the point between $0$ and $m-1$. For a hash function to be considered a good minhash we require the following conditions:

\begin{enumerate}
\item The hash function needs to be easy to compute.
\item It should be easy to extend the hash function to a family of hash functions allowing an arbitrary number of minhashes to be created for the same point.
\item The probability of a collision needs to be related to the distance between the points and it has to be a continuous monotonous function. So if $d(x,y) \leq d(x,z)$ then $P[h(x)=h(y)] \geq P[h(x)=h(z)]$. The inequalities are not relevant as long as the function is monotonous and continuous.
\end{enumerate}

We say an LSH family of hash functions $\mathcal{H}$ is $p_1,p_2$ sensitive for $d_1$ and $d_2$ distances: $\mathcal{H}(d_1,d_2,p_1,p_2)$ if for any two given points $x,y$ then when the points are at distance $d_1$ or less then $P[h(x)=h(y)] \geq p_1$ and when the points are at a distance $p_2$ or greater then $P[h(x)=h(y)] \leq p_2$. 

It can be shown that conditions 2 and 3 are enough to amplify any family $\mathcal{H}$ to arbitrary values of $p_1$ and $p_2$, this can be done using a process known as \textit{amplification}.

\section{Amplificacion of LSH Families}

For any LSH family $\mathcal{H}(d_1,d_2,p_1,p_2)$ where $p_1$ is the probability of retrieving points that are close to a query point, we would like $p_1$ to be as high as possible. $1-p_1$ is the occurrence of \textit{false negatives} meaning that some points that are closer than $d_1$ to our query point won't be retrieved. On the other hand $p_2$ is the probability of retrieving points that are further than desired to our query point; $p_2$ is the probability of \textit{false positives} we would like $p_2$ to be as small as possible. In general terms false negatives affect the precision of the LSH scheme while false positives affect its performance as the number of distances that need to be compared can be high. 

The rate of false positives can be reduced using more than one minhash over the same hash table, if we use $r$ minhashes then the candidate points are the intersection of the points found in the buckets pointed by $h_{i=0}^{r-1}(x)$. This applies to both $p_1$ and $p_2$ so in general for $r$ hash functions used in conjunction the collision probability $p$ becomes $p^r$. 

The rate of false negatives can be reduced using more than one hash table, then we can say that a point is candidate to be close to our query if it is a candidate in either of the $b$ hash tables used. This means that the probability of collisions $p$ is $1-(1-p^r)^b$ 

So using $r$ hash functions and $b$ hash tables we get the LSH family $\mathcal{H}(d_1,d_2,1-(1-p_1^r)^b,1-(1-p_2^r)^b)$

As long as the probability of a collision is a monotonous function of the distance between the points then, for any LSH family, we can achieve arbitrary values for $p_1$ and $p_2$. The cost of increasing $b$ is related to space as more hash tables are needed while the cost of increasing $r$ is related to performance as more hash functions are needed. The process of amplification shows why condition 2 is important as an arbitrary number of different minhashes might be needed.

A minhash function is neutral if it's probability of collision is equal to 1 minus the normalized euclidean distance between the points. Minhash functions that show a curve above this line have a higher probability of collision so they have less false negatives and more false positives. Minhash functions below the neutral line have a lower collision probability and then produce less false positives but more false negatives. Depending on the constraints of space or performance we might prefer one case or the other. This is an important observation because it directly points to a practical rule for choosing minhash functions.

\section{Previous Work}

\subsection{Hyperplanes LSH}

Our first LSH family for the angular distance is based on random \textit{hyperplanes} minhashes and was proposed by Charikar in 2002 [4]. The construction is very simple: every minhash uses a random hyperplane $v_i$ of the same dimensionality as the points to be hashed, the minhash is then defined as the sign of the dot product between the point and the random vector. $h_i(x) = \textit{sign}<v_i,x>$. Since the minhash can only take 2 values each minhash defines a bit, we can create a $d'-bit$ minhash using $d'$ minhashes. Since each hyperplane can be seen as a partition of the unit hypersphere in two halves then we can see that the probability of collision is related to the distance \footnote{from now on we'll use distance to refer indistinctly to the angular distance or the euclidean distance in the unit hypersphere} between the points. In concrete the probability of collision is $1-\frac{\alpha}{\pi}$ where $\alpha$ is the angle between the vectors.

It was also shown in [4] that a random vector formed by just $\pm1$ elements is enough as a random hyperplane. This is related to a Johnson Lindenstrauss projection presentd by Achlioptas[1].

\subsection{Voronoi LSH}

Voronoi LSH [3] uses $T$ random Gaussian vectors of the same dimensionality as the data points. The minhash is defined as $h(x) = \argmax_{i=0...T-1}{<G_i,x>}$ This means that using $T$ Gaussian vectors each minhash can create $T$ different values. It is easy to show that if $T=2$ then Voronoi LSH is the same as Hyperplanes LSH because choosing the closest point from two random points in the sphere is the same as randomly partitioning the sphere bisecting the two random points. In general Voronoi LSH with $T=2^K$ Gaussian vectors is similar to Hyperplanes LSH with $k$ hyperplanes. 

\subsection{Cross Polytope LSH}

The Cross Polytope LSH method was introduced by Teresawa and Tanaka in 2007 [10]. Each minhash uses a random rotation from $d$ to $d'$ dimensions and then the nearest vertex of the d'-dimensional cross polytope is chosen as the value of the minhash. In $d'$ dimensions each polytope has $2d'$ vertices, for example in $\Re^2$ the polytope is a square determined by the vertices (0,1);(1,0);(-1,0) and (0,-1). 

The Cross Polytope method is actually a variant of Voronoi Hashing, if we accept a Gaussian matrix as a pseudo-rotation and we consider the maximum absolute value plus the sign of the element instead of just the maximum value then Voronoi Hashing is the same as the Cross Polytope method. For example using 5 Gaussian vectors we might get the following results for two points $x,y$: $x \rightarrow (3,2,-5,-1,2)$ $y \rightarrow (1,4,-6,3,1)$. Using Voronoi Hashing the minhash for $x$ is 0 the index of the maximum value while the minhash for $y$ is 1. Using Cross Polytope we observe that both would be closer to the vertex $(0,0,-1,0,0)$ and this is the same as taking the index and sign of the maximum absolute value of the vectors as the minhash which is 2 for both vectors. 

To speed up the computation of the rotation a pseudo-rotation using Hadamard matrices has been proposed [2].

\subsection{Fast Cross Polytope LSH}

Kennedy[8] proposes a faster version of the Cross Polytope method using a Fast Johnson Lindenstrauss transform from the original $d$ dimensions to a reduced space with $m$ dimensions and then the random lifted rotation from $m$ to $d'$ dimensions. Our experiments show that for small dimensionality vectors this method is actually slower in practice that a direct random rotation and the other methods studied. As the dimensionality of the vectors is larger this method can become more efficient but then a direct dimensionality reduction of the vectors using feature hashing can be applied.

\subsection{Even Faster Cross Polytope LSH}

Instead of a FJL transform feature hashing can be used making the method from [8] a lot faster. This means FH is used to project from $d$ to $m$ dimensions and then a random rotation is used from $m$ to $d'$ dimensions. Experiments show that the results offer very similar precision but are significant faster but yet not as fast as the other methods studied in this work.In general the two-step approach from $d$ to $m$ and from $m$ to $d'$ to obtain a minhash is not a speed improvement over a direct minhash from $d$ to $d'$.

\section{Johnson-Lindenstrauss Projections for LSH}

We have mentioned that any function $h(x)$ where $P[h(x)=h(y)] = f(||x-y||)$ with $f$ monotonous can be used as a minhash. In particular a family of functions that transforms the points from one dimensional space to another preserving the norms of the vectors can be used and thus any projection derived from the Johnson-Lindenstrauss lemma is usable as a minhash. This can be proven for sparse vectors independently of the number of vectors to be used via the Restricted Isometry Property (RIP) as described in [16], for non-sparse vectors the number of points matters as the Johnson-Lindenstrauss is applied and then for a large number of points the pairwise distances between vectors is preserved with a small error. We notice that if the number of points is not small then we don't need to use LSH so we can say that the pairwise distances between points are preserved with a high probability.

Voronoi LSH is indeed a direct application of a Johnson-Lindenstrauss projection using a Gaussian Matrix [1]. Hyperplanes LSH is another application of a JL projection using a matrix with $\pm1$ random elements [1].

We now show that for any random projection that is also a Johnson-Lindenstrauss transform two generic families of LSH functions can be created defining two minhashes.

\begin{theorem}
If $p$ is a randomizable Johnson-Lindenstrauss projection from $d$ to $d'$ dimensions then $p$ can be used to construct at least two different LSH families for the angular distance based in the following minhashes.

\begin{enumerate}
\item $\argmax_{0..d'-1}{<p_i,x>}$ with $p_i$ being the $i$th column of $p$
\item $\text{sign}({<p_i,x>})$ with $p_i$ being the $i$th column of $p$
\end{enumerate}
\end{theorem}

The theorem is almost self-proven. For method 1 we are comparing the closest point from a set of random points in the unit sphere, since the projection preserves the distances the probability of a collision is a monotonous function of the original distance between the points. For method 2 we are bisecting the sphere in two halves, the probability of two points being in the same halve is again a monotonous function of the original distance between the points because $p$ preserves the distances between the points.

It can be seen that when $p$ is a Gaussian projection method 1 is Voronoi Hashing. And when $p$ is a random projection filled with $\pm1$ method 2 is hyperplanes LSH.Remembering that Cross-Polytope LSH is a form of Voronoi LSH this means that all the LSH families we have described can be generalized to the creation of a minhash from a Johnson-Lindenstrauss projection.

In this work we will show that Feature Hashing is also a JL projection and that will be the theoretical foundation to use it to create a minhash and a LSH family for the angular distance.

\section{Feature Hashing for LSH}

Feature Hashing [9] , also known as \textit{The Hashing Trick} is a very simple method for dimensionality reduction from the original $d$ dimensions to $d'$. A nice advantage is that when a feature of our data is categorical it can be hashed into several dimensions without needing to know the total number of different values for the feature. To mitigate the effect of collisions Weinberger [11] proposed the use of a second hash function that will return the sign to be used ($\pm1$). With the addition of the second function the effect of collisions is mitigated and several features can be hashed into the same target space with minimal interplay[11]. 

\subsection{Feature Hashing is a Johnson Lindenstrauss Transform}

Feature hashing as a random projection is used in [16] to transform sparse vectors preserving pairwise distances, [11] also shows that FH preserves the distances between vectors. 

We start showing that Feature Hashing can be represented as a matrix. Each feature (column) of our original vectors is mapped by the first hash function from $d$ dimensions to $d'$ and the second hash function determines the sign. This is the same as multiplying the original ($1xd$) vector by a $dxd'$ matrix where each row contains exactly one $\pm1$ element. If we use $k$ hash functions instead of just one then each row contains up to $k$ $\pm1$ elements and it might contain up to $\pm k$ valued elements due to collisions. 

So for example if we have the data-point $V=(0,1,0,3,0.5,0,1)$ in $\Re^7$ and we use FH to convert it to $d'=4$ dimensions we can use the following hash function:

\begin{verbatim}
h(0) = 2   s(0)=+1
h(1) = 1   s(1)=+1
h(2) = 3   s(2)=-1
h(3) = 0   s(3)=+1
h(4) = 1   s(4)=-1
h(5) = 2   s(5)=-1
h(6) = 3   s(6)=-1
\end{verbatim}

This means the resulting vector is $V'=(3,0.5,0,-1)$. 

This is the same as multiplying our $1x7$ vector by a $7x4$ matrix in the form:

$$
\begin{pmatrix}
0&1&0&3&0.5&0&1\\
\end{pmatrix}
*
\begin{pmatrix}
0&0&1&0\\
0&1&0&0\\
0&0&0&-1\\
1&0&0&0\\
0&-1&0&0\\
0&0&-1&0\\
0&0&0&-1\\
\end{pmatrix}
=
\begin{pmatrix}
3&0.5&0&-1\\
\end{pmatrix}
$$

Now we'll show that these kind of matrices preserve the norms of the vectors from the original space which is a generalization of [9] and [17].

\begin{theorem}
Given $A$ a $dxd'$ matrix where each row has $k$ $\pm1$ elements and the rest are zeros. We want to show that this matrix preserves the norms of the vectors from $d$ dimensions projected into $d'$ dimensions up to a given scale factor.

$$
||w|| \approx \sqrt{k} ||v||
$$
\end{theorem}

To prove the theorem we start from the fact that if we have a vector $v_d$ in $\Re^d$ and we apply a dense random projection with $\pm1$ elements the square of the projection has an expected value equal to the square of the norm this is because of concentration of measures.

Each column of our $dxd'$ matrix is equivalent to a random projection of $d$ elements, hence when we compute $v * A$ we are obtaining $d'$ different random projections. The difference is that our matrix doesn't have $d$ $\pm1$ elements in each column. The total number of non-zero elements in our matrix is $d*k$ ignoring collisions. Then the average number of non-zero elements in our matrix is $\frac{dk}{d'}$

Since we know that with $d$ $\pm1$ elements the square of the projection has an expected value equal to the square of the norm, then if we have $\frac{k}{m}$ $\pm1$ elements the square of the projection has an expected value equal to $\frac{k}{m}||v||^2$

$$
w_i^{2} = \frac{k}{m}||v||^2
$$

We can now compute the expected value of the norm of $v_m$ as:

$$
\begin{aligned}
w_i^2 = \frac{k}{m}||v||^2\\
\sum_i w_i^2 = m\frac{k}{m}||v||^2\\
\sqrt{\sum_i w_i^2} = \sqrt{m\frac{k}{m}||v||^2} \\
||w|| = \sqrt{k} ||v||\\
\end{aligned}
$$

And this proves the theorem. 

\subsection{Method 1: Feature Hashing LSH}

Having shown that Feature Hashing is a form of Johnson-Lindenstrauss projection and established that any Johnson-Lindenstrauss projection can be used as a minhash, we propose two LSH families based in Feature Hashing. The first one is a direct application of Feature Hashing.

Each minhash uses a random $dxd'$ matrix $M$ with $k$ $\pm1$ elements in each row. The minhash is then computed as:

$$
h(x) = \argmax_{i=0..d'-1} <x,M_i> 
$$

Where $M_i$ is the $i$th column of $M$.
Using different random matrices we can create different minhashes to amplify the family.

\subsection{Method 2: Directional Feature Hashing LSH}

In method 2 we are again using a $dxd'$ matrix $M$ with $k$ $\pm1$ elements, then we consider the sign of each element in $d'$ and build a $d'$-bits minhash as the result. This is the same as hyperplane LSH but using only $k$ $\pm1$ elements in each random hyperplane. If the method is usable then we have a direct improvement in performance compared to the LSH families we have studied.

$$
h(x) = \text{sign}(<x,M_i>)
$$

Where $M_i$ is the $i$th column of $M$

\section{Results}

We now turn to results we obtained applying these methods to a synthetic and real dataset. For the synthetic dataset we build random vectors with $d=128$ dimensions in the unit hypersphere. For the real dataset we used the SIFT 1 million dataset [5] that is widely used for Near Neighbors experiments. We note that for SIFT the vectors are not in the Unit Hypersphere but the methods proposed work very well for the euclidean distance showing that in the practice these LSH families can be used for both angular or euclidean distances.

\begin{figure}[htp]
\centering
\fbox{\includegraphics[width=3.5in]{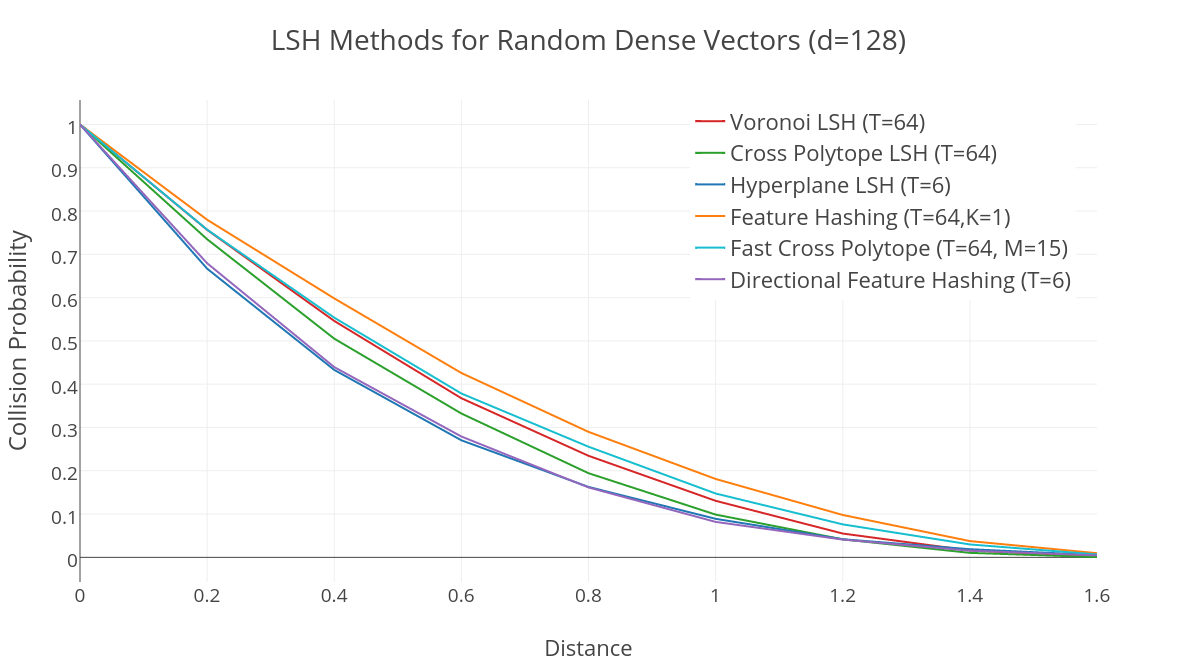}}
\caption{Results for random vectors in the unit hypersphere for d=128}
\label{dense128}
\end{figure}

The first graph (Figure~\ref{dense128}) shows the probability of collisions by the Euclidean distance for different LSH families in the Unit hypersphere.

The graph shows several interesting things. Hyperplane LSH is almost the same as Directional Feature Hashing but the second is faster because its vectors are sparser. Both families have the lowest rate of false positives but also the highest rate of false negatives. Using less bits for each minhash it is easy to decrease the number of false negatives with the cost of increasing the false positives.

Feature Hashing has the highest rate of false positives and the lowest rate of false negatives. The effect can be mitigated using more hash functions for the same minhash, which means using more than 1 $\pm1$ element in each row of the matrix. Voronoi LSH, Cross Polytope and the Fast Cross Polytope are very similar, which is expected as we have shown they are almost exactly the same thing. 

For the SIFT dataset we run an experiment setting a desired value of $p=0.95$ for distances 0.2 or less and $p=0.05$ for distances 0.6 or more. This means that vectors that are at distance 0.2 or less will have a probability of collision of 0.95 or more while vectors that are at distances 0.6 or more will have a probability of collision of 0.05 or less.

The idea was to see which values of $r$ (number of hash functions) and $b$ (number of hash tables) were needed to amplify each minhash to the target probabilities for each method.

\begin{table}[!hbt]
\begin{center}
\begin{tabular}{|l|c|c|c|}
\hline
	\textbf{Method}& \textbf{r} & \textbf{b} & \textbf{Total}\\
\hline
	Voronoi LSH (T=64) & 6 & 15 & 90\\
\hline
	Cross Polytope (T=64) & 6 & 18 & 108\\
\hline
	Hyperplane (T=6) & 5 & 22 & 110\\
\hline
	Feature Hashing (T=64,k=1) & 7 & 16 & 112\\
\hline
	Fast Cross Polytope (T=64)& 6 & 15 & 90\\
\hline
	Directional Feature Hashing (T=6) & 5 & 20 & 100\\
\hline
\end{tabular}
\end{center}
\caption{Number of hash functions and hash tables needed for $p1=0.95$ and $p2=0.05$}
\end{table}

The results are very interesting, we can see that all the methods use a similar number of total hash functions. Hyperplane LSH and Directional Feature Hashing need less functions per table but more tables, so they are very good when space is not limited and performance is critical. Feature Hashing needs less hash tables but more hash functions so it is a family to consider when space is critical. In general each LSH family can work better or worst depending on the data and the target probabilities for false negatives and false positives. All these LSH families are usable and need to be considered carefully when choosing a LSH family for an application.

Next we did some speed tests on the SIFT dataset, we multiplied the speed of a single hash function by the number of hash functions needed in each family to obtain a 0.05 probability for both false positives and false negatives. This is done to even out the advantages and disadvantages of each particular function.

\begin{figure}[htp]
\centering
\fbox{\includegraphics[width=3.5in]{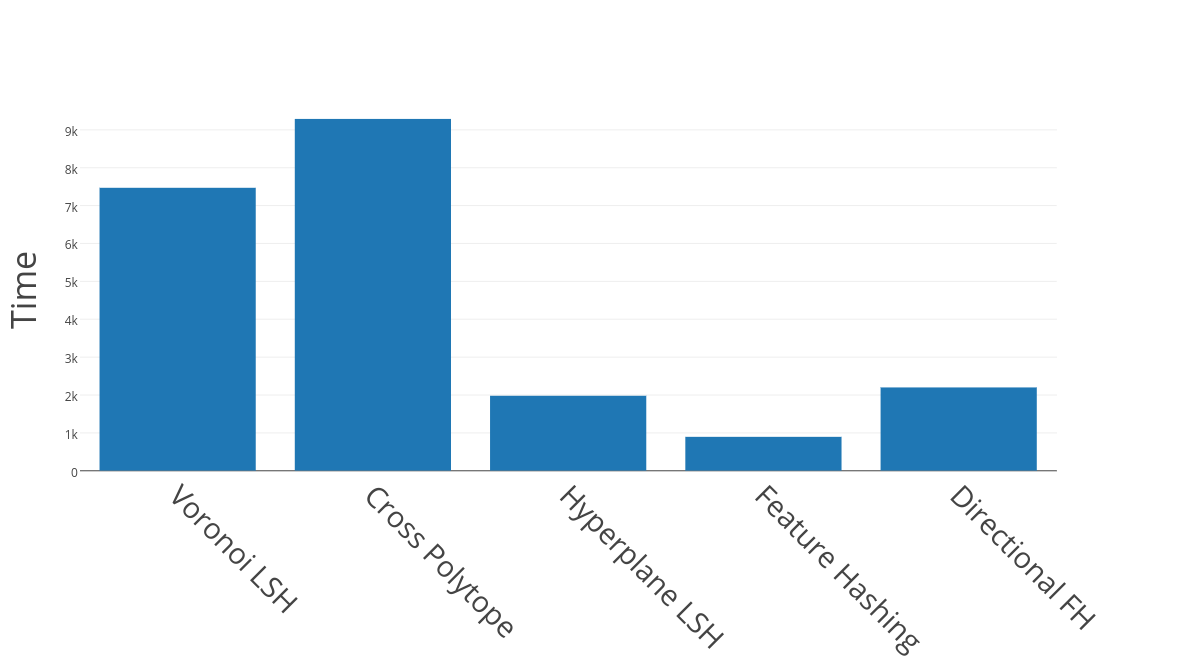}}
\caption{Performance of LSH methods}
\label{speed}
\end{figure}

The Fast Cross Polytope was discarded because it was very slow compared to the other methods, this is because the FJL transformation takes time and then a further rotation is needed. The two step approach was never faster than the direct application of a simpler LSH family. Cross Polytope and Voronoi are very similar because they are the same thing. Hyperplane Hashing and Directional Feature Hashing are much faster and the direct application of Feature Hashing was the fastest method. This means that even needing more hash functions to mitigate false positives Feature Hashing is still the fastest LSH family for the SIFT dataset. This is very logical as each hash function only does a limited number of additions and subtractions, no multiplications are needed and each element in the vector is only added or subtracted once to the resulting vector.

The rate of how precision changes was also studied.

\begin{figure}[htp]
\centering
\fbox{\includegraphics[width=3.5in]{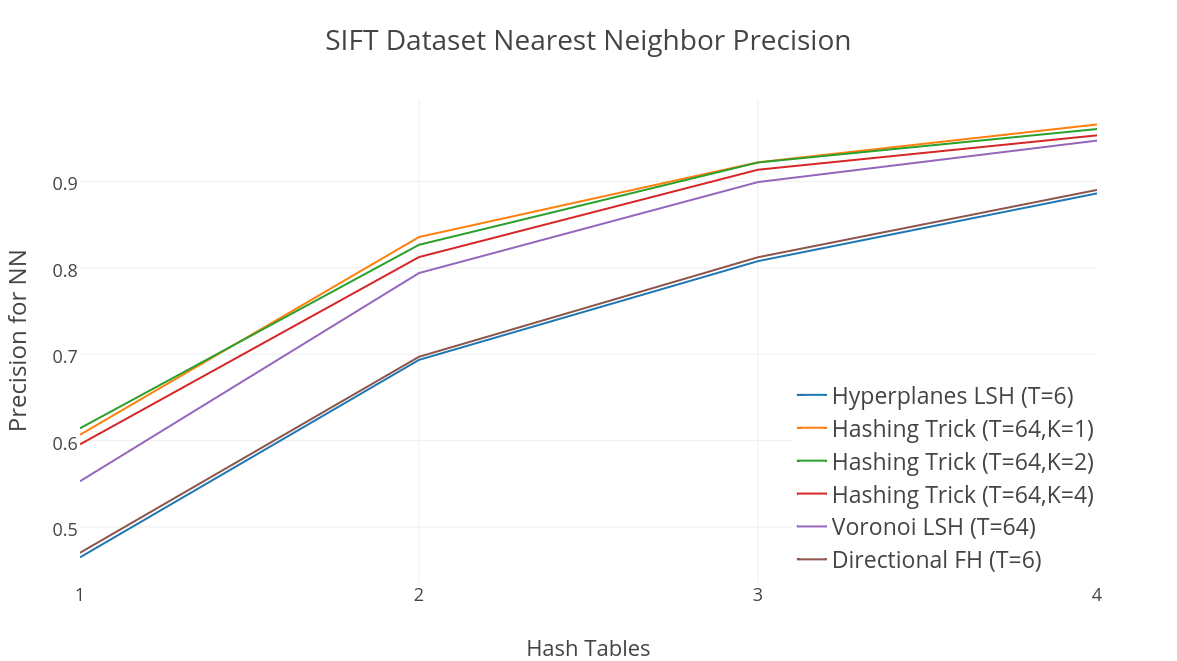}}
\caption{Precision rate as more Hash tables are used}
\label{rate}
\end{figure}

It can be seen (Figure~\ref{rate}) that as more hash tables are used the precision increases in an almost identical rate in all the LSH families tested, this is logically explained as they are all based on a form of a Johnson-Lindenstrauss transform that preserves the norms of the vectors. The study was made to discard the possibility of an LSH family having a steeper precision increase which would mean that less functions would be needed to achieve a target precision compared to the other methods. As it can be seen that is not the case. 

\section{Analysis}

\subsection{Feature Hashing LSH}

Feature Hashing is a very flexible LSH family. It can be applied in matrix form or using hash functions, the later is very practical for data with categorical columns or text where the other methods can't be used without doing a data transformation first.  In matrix form we have only $k$ non-zero values in each row making its time complexity  $O(d*k)$. Since each member of the original vector is only used once and only an addition or subtraction is performed, we can claim that the method is optimal in terms of speed.

The number of hash functions or the number of non-zero values in each row of the associated matrix can be tuned to reduce the number of false-positives independently of amplification. Adding more $\pm1$ elements decreases the rate of false positives at the cost of increasing the computation for each minhash.

\begin{figure}[htp]
\centering
\fbox{\includegraphics[width=3.5in]{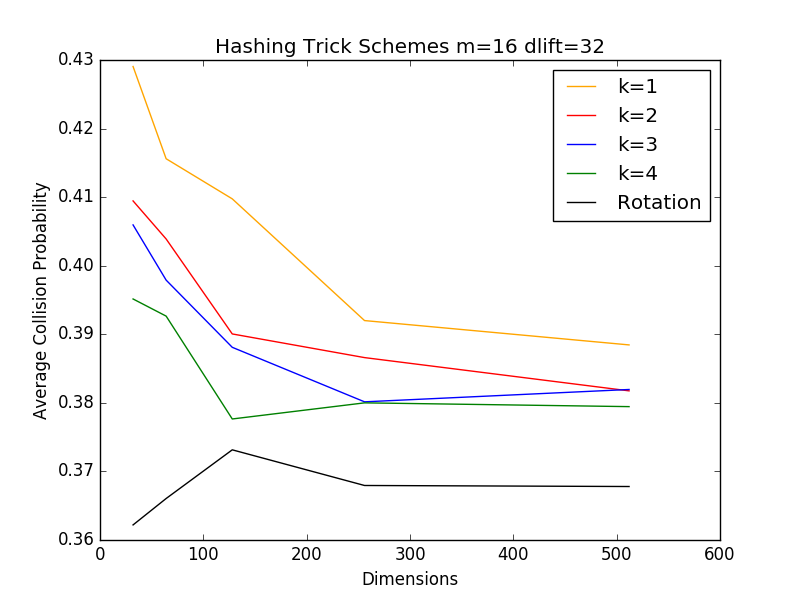}}
\caption{Collision Rate as More hash functions are used}
\label{morek}
\end{figure}

We observe (Figure~\ref{morek}) that for vectors in 128 dimensions we don't need to use a full 128 dimension rotation to reduce the collision probability and the number of false positives. Even using only a few $\pm1$ elements is enough to decrease the rate of false positives. Sensitivity tests are needed to evaluate the optimal number of $\pm1$ elements in terms of performance and false-positive rates.

The most interesting result is that for a desired LSH family in the form $\mathcal{H}(d_1,d_2,p_1,p_2)$ Feature Hashing is the fastest method at exactly the same rate of false positives and false negatives.

\subsection{Directional Feature Hashing LSH}

Directional Feature Hashing is very similar to Hyperplanes LSH but faster because only $k \pm1$ elements are present in each projection. When $k=d$ the method is equal to Hyperplanes Hashing. The number of bits can be tuned to make the method work as expected independently of the amplification used.

This LSH family shows that hyperplanes LSH can be made faster just sparsifying the random projections used.

\section{Conclusion}

We have generalized LSH families for the angular distance showing the requirements for a function to be considered a minhash. In general terms any randomized function that depends on the distance between the vectors is suitable as a minhash function. The number of different minhashes that can be created is the size of the LSH family and a high number is desired.

We showed how any minhash can be extended to work within the parameters of false positives and false negatives we expect using amplification. We proved that any form of a random Johnson-Lindenstrauss projection can be used to create an LSH family for the angular distance because the projections preserve the norms of the vectors. 

We showed how Feature Hashing is a form of a Johnson-Lindenstrauss projection. Then two new LSH families were proposed based in Feature Hashing, one with a low rate of false negatives and a higher rate of false positives and the other with a low rate of false positives and a higher rate of false negatives. Depending on the constraints in performance and space one or the other can be used and amplified to achieve the desired results. A very important characteristic of the two methods presented is that they are very fast in performance and very simple to implement. 

In terms of optimality Spherical hashing [3] is optimal in terms of precision but the minhashes are not practical to compute while Feature Hashing as presented here is optimal in terms of speed and the corresponding minhashes can be used in the practice because they offer good precision after amplification.





%

\end{document}